\documentclass{raa}            
\usepackage{graphicx,times}
\usepackage{natbib}

\providecommand{\bm}[1]{\mbox{\boldmath$#1$\unboldmath}}

\begin{document}

   \title{On the non-evolution of the dependence of black hole masses on bolometric luminosities for QSOs
}

 \volnopage{ {\bf 2012} Vol.\ {\bf 12} No. {\bf XX}, 000--000}
   \setcounter{page}{1}

   \author{M. L\'opez-Corredoira\inst{1,2} and C. M. Guti\'errez\inst{1,2}}

   \institute{$^1$ Instituto de Astrof\'\i sica de Canarias,
E-38200 La Laguna, Tenerife, Spain\\
$^2$ Departamento de Astrof\'\i sica, Universidad de La Laguna,
E-38206 La Laguna, Tenerife, Spain\\
{\it martinlc@iac.es}\\
\vs \no
   {\small Received 2011 August 3; accepted 2011 November 19}
}

\abstract{ There are extremely luminous quasi stellar objects (QSOs) at high redshift which are absent at low redshift. 
The lower luminosities at low redshifts can be understood
as the external manifestation of either a lower Eddington ratio or a lower mass. 
To distinguish between both effects, we determine the possible dependence of masses and Eddington ratios
of QSOs with a fixed luminosity as a function of redshifts; this avoids
the Malmquist bias or any other selection effect.
For the masses and Eddington
ratios derived for a sample of QSOs in the Sloan Digital Sky
Survey, we model their evolution by a double linear fit separating the dependence on
redshifts and luminosities. The validity of the fits and possible systematic
effects were tested by the use of different estimators of masses or bolometric
luminosities, and possible intergalactic extinction effects.
The results do not show any significant evolution of black hole masses or Eddington ratios for equal
luminosity QSOs. The black hole mass only depends on the bolometric luminosity without significant dependence on the redshift
as \protect{$\left(\frac{M_{BH}}{10^9\ M_\odot}\right)\approx 3.4\left(\frac{L_{bol}}{10^{47}\ {\rm erg/s}}\right)^{0.65}$} on average for \protect{$z\le 5$}.
This must not be confused with the possible evolution in the formation of black holes in QSOs. 
The variations of environment might influence the formation of the black holes but not its subsequent accretion. 
It also leaves a question to be solved: Why are there not QSOs with very high mass at low redshift?
A brief discussion of the possible reasons for this is tentatively pointed out.
\keywords{accretion  --methods: statistical--- quasars: general}
}

\authorrunning{L\'opez-Corredoira \& Guti\'errez}            
\titlerunning{Non-evolution QSOs}  
\maketitle


%
%

\section{Introduction}
\label{.introd}

Quasi stellar objects (QSOs) are extremely bright at high redshift, but at low redshift they are much less luminous; in fact from the analysis of the bolometric luminosity function of
QSOs at different redshift (Hopkins et al. 2007), it is clear that the relative
abundance of high luminosity QSOs decreases quickly at low redshift.  In the Sloan Digital Sky Survey (SDSS), all QSOs at $z<0.4$ are fainter than $M_B=-26$  (with K-corrections) while there are numerous QSOs tens of times brighter than this limit at higher
redshifts. The same effect is also observed in other spectral ranges; for instance in 
X-rays  (Shen et al. 2006) or in radio  (Bridle \& Perley 1984; Bell 2006, Figs. 9,
10). All these observations require a strong density and luminosity evolution. 

Something very different must have happened at high redshift with respect to the low
redshift Universe to obtain this different level of luminosity. However, no visible
signs of this evolution are observed. There is no indication of any significant
evolution in the X-ray properties of  quasars between redshifts 0 and 6, apart from
the intrinsic luminosity,  suggesting that the physical processes of accretion onto
massive black  holes have not changed over the bulk of cosmic  time (Vignali et al.
2005). Also, the spectral features of low and high  redshift QSOs are very similar
(Segal \& Nicoll 1998). This strong change in luminosity without any additional
external sign of evolution is one of the most relevant pending problems in QSOs 
(L\'opez-Corredoira 2011).

There are two possible reasons for the lower luminosity of low $z$ QSOs with respect
to those ones with high $z$: 1) either their black holes are less massive, and this
would explain the lower power of their accretion disks, 2) or they have obtained approximately the same mass but they are less efficient, due to a lower accretion
rate. There are several works that estimated masses and Eddington ratios of QSOs' black holes (McLure \& Dunlop 2004, Kollmeier et al. 2006, Vestergaard \& Peterson
2006, La Mura et al. 2007, Vestergaard et al. 2008, Shen et al. 2008 [herein S08],
Labita et al. 2009a,b). The result in general is that masses and Eddington ratios are
larger at higher $z$. 
Nonetheless, it cannot be directly interpreted as a sign of
evolution of QSOs because the samples used are strongly affected by the Malmquist bias,
that is, we are comparing low luminous QSOs at low $z$ with high luminous QSOs at high
$z$. S08 made a separation in bins of fixed luminosity showing qualitatively that much
of this apparent evolution is due to luminosity variation. We think that this analysis
requires further attention in order to properly quantify it and the possible
systematics involved. This is the main aim of this paper in which we develop a simple
method to study the dependence of black hole masses and Eddington ratios on their
redshifts and luminosities.

\section{Sample and methodology}
\label{.samplemethod}

A complete and exhaustive compilation of masses and bolometric luminosities of
QSOs from SDSS is provided by S08, who determined the masses from the
continuum and width of
the broad emission lines of H$_\beta$, MgII-2798\AA \ or CIV-1549\AA \ in the ranges
$z<0.7$, $0.7<z<1.9$ and $1.9<z<5.0$ respectively. The total number of objects
cataloged by S08 is 77,429. Shen et al. (2011) contains that information for 105,783 QSOs obtained
from a more recent version of SDSS, but we still use S08 since we have carried out our analyses with this sample, 
and we do not expect any significant improvement in our results by just adding 30-40\% more QSOs.  
From this catalog we selected those objects for which there are 
reliable estimation of masses and luminosities. We imposed some additional constraints based 
on the width of the lines ($FWHM>2000$ Km/s) and the SNR ($>5$) of the continuum.
So, the sample selected has 49,213 objects with an average redshift of $\langle z\rangle =1.35$. From the black hole masses, $M_{BH}$, 
and the bolometric luminosities, $L_{bol.}$, it is straight forward to determine 
the Eddington ratio 
\begin{equation}
\epsilon =\frac{L_{bol.}}{L_{Edd.}}
,\end{equation}
where $L_{Edd.}$ is the Eddington luminosity
(given by e.g., Kembhavi \& Narlikar 1999, eq. (5.26)):
\begin{equation}
L_{Edd.}=1.3\times 10^{38}\left(\frac{M_{BH}}{M_\odot}\right) \ {\rm erg/s}
\end{equation}

In order to avoid the Malmquist bias, we separate the dependence of the interesting
quantities on redshifts and on luminosities in the following way:
given a variable $r$, dependent on the redshift ($z$) 
and the absolute magnitude with K-correction ($M_{i,rest}$, AB-magnitude) 
as independent variables, 
we model the dependence of such variables as 
a bi-linear function of redshift and luminosity
\begin{equation}
r=a+bx+cy
,\end{equation} 
\[x\equiv f(z); y\equiv g(M_i) 
.\]
We use $f(z)=z$, $g(M_i)=M_{i}+23$.
The double linear fit of the points $r_i(x_i,y_i)$ gives us the values of $a$, $b$ and
$c$; i. e. in this way we separate the dependence on redshift and luminosity. The
coefficients $b$ and $c$ are easy to interpret as the ratio of evolution with redshift
and luminosity respectively. The model is extremely simple and in principle the
dependence might contain non-linear terms, but in any case, $b$ and $c$ will reflect
the ``average'' gradients in the dependence on $z$ and $M_i$. 
We do not say that given a luminosity and a redshift there will be   
only one possible value of $r$; we do not claim that the 
deviation of $r$ from luminosity is due
to redshift alone o vice verse. However, we can talk about an   
average dependence on $z$ and $M_i$.
Given a set of QSOs with the same luminosity and the   
same redshift, we will measure an average mass and an r.m.s. value due to   
both the possible dependence on other variables and the errors in  
the measurements. On the other hand, the
analysis of the residuals between the data and our simple bi-linear model indicates that
the fit is a good description of the data (see below).

\begin{figure}
\vspace{1cm}
{\par\centering \resizebox*{7cm}{7cm}
{\includegraphics{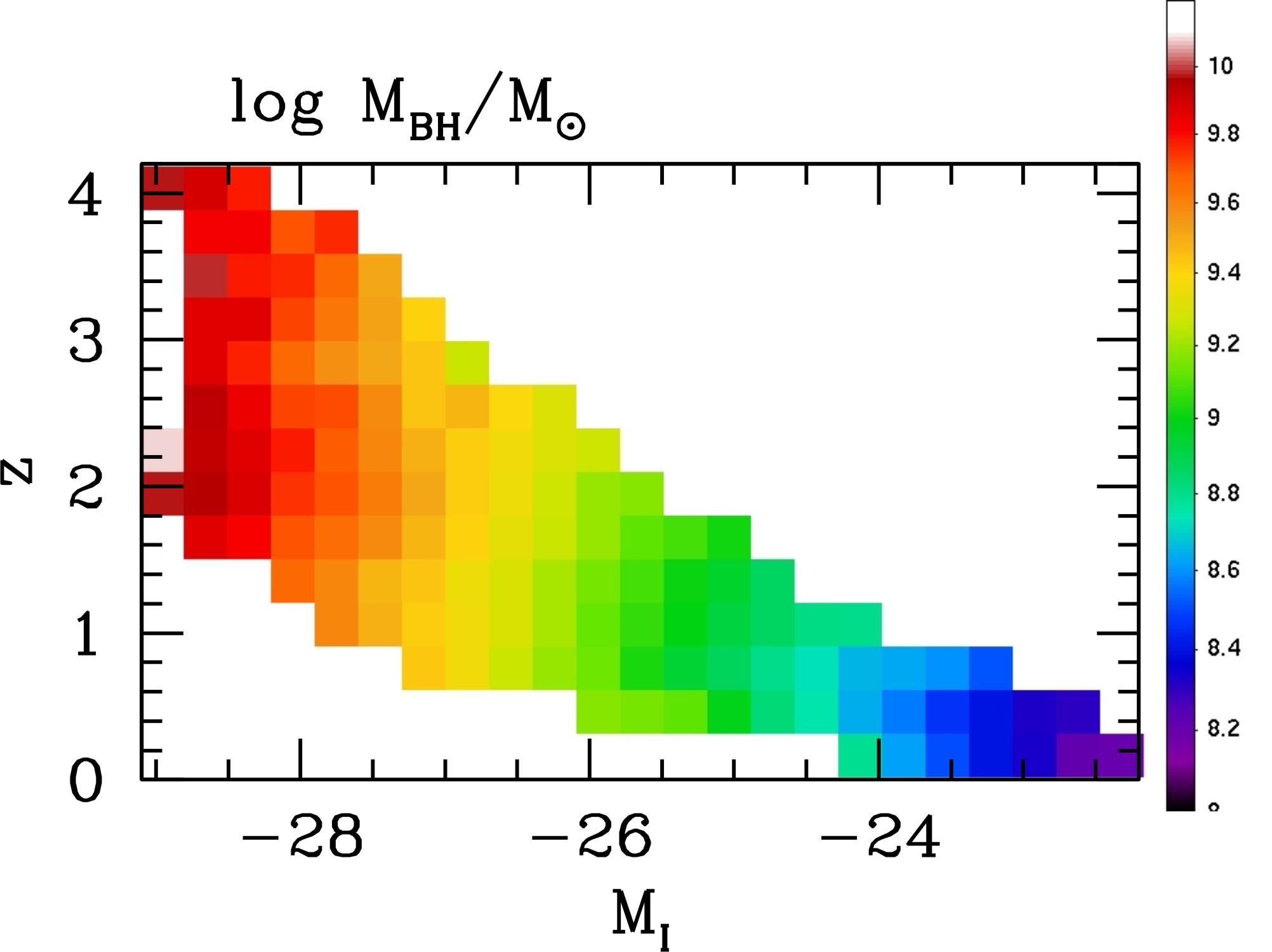}}
\par\centering \resizebox*{7cm}{7cm}
{\includegraphics{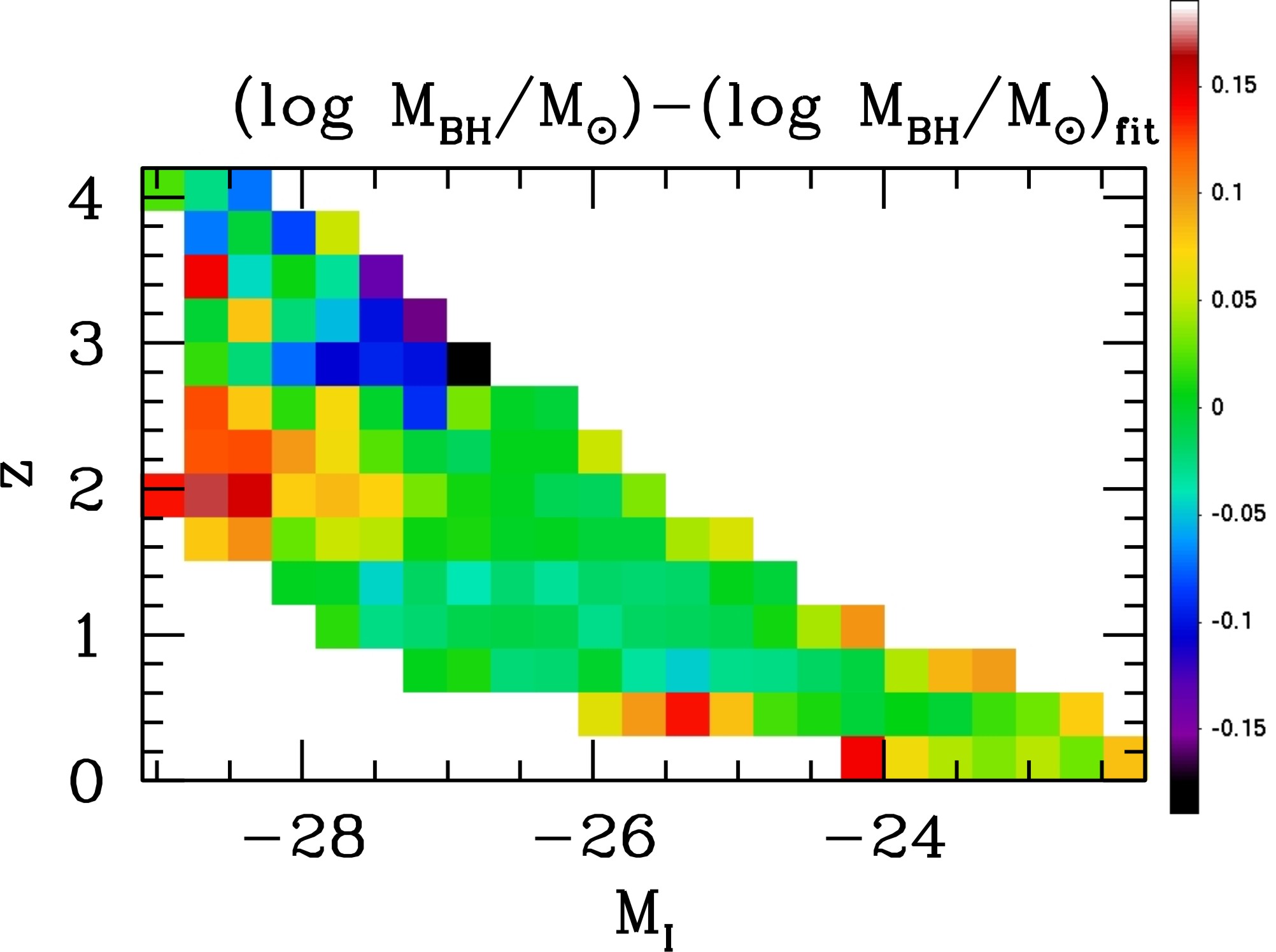}}
\par \centering }
\caption{($Top$) Average of $\log _{10}M_{BH}$ (logarithm of the black hole mass)
in bins of $\Delta z=0.3$ in redshift and
$\Delta M_i=0.3$ in the $i$-rest absolute magnitude. ($Bottom$) Difference between the data and
the best fit model.}
\label{Fig:mass}
\end{figure}

\begin{figure}
\vspace{1cm}
{\par\centering \resizebox*{7cm}{7cm}{\includegraphics{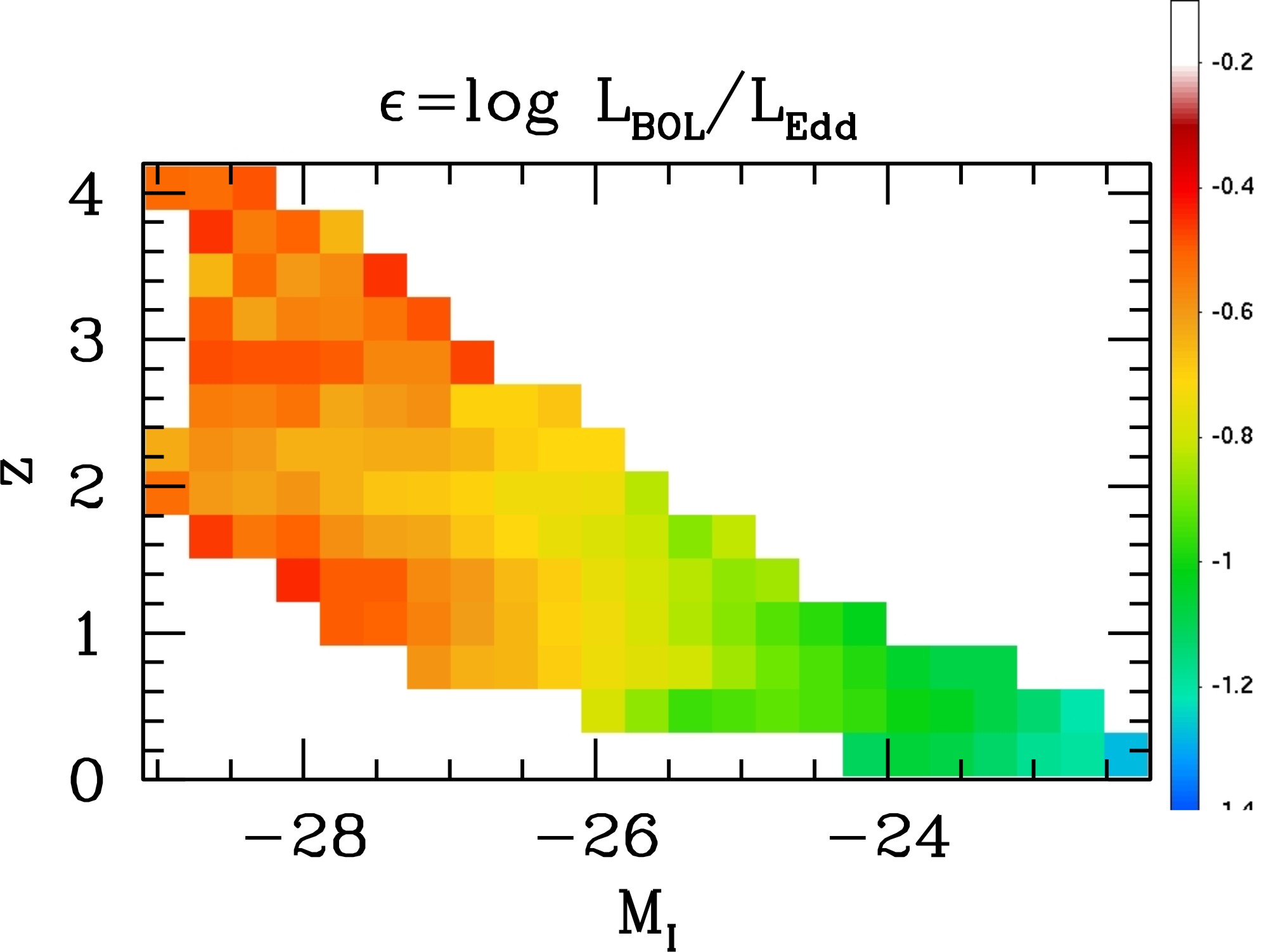}}
\par\centering \resizebox*{7cm}{7cm}{\includegraphics{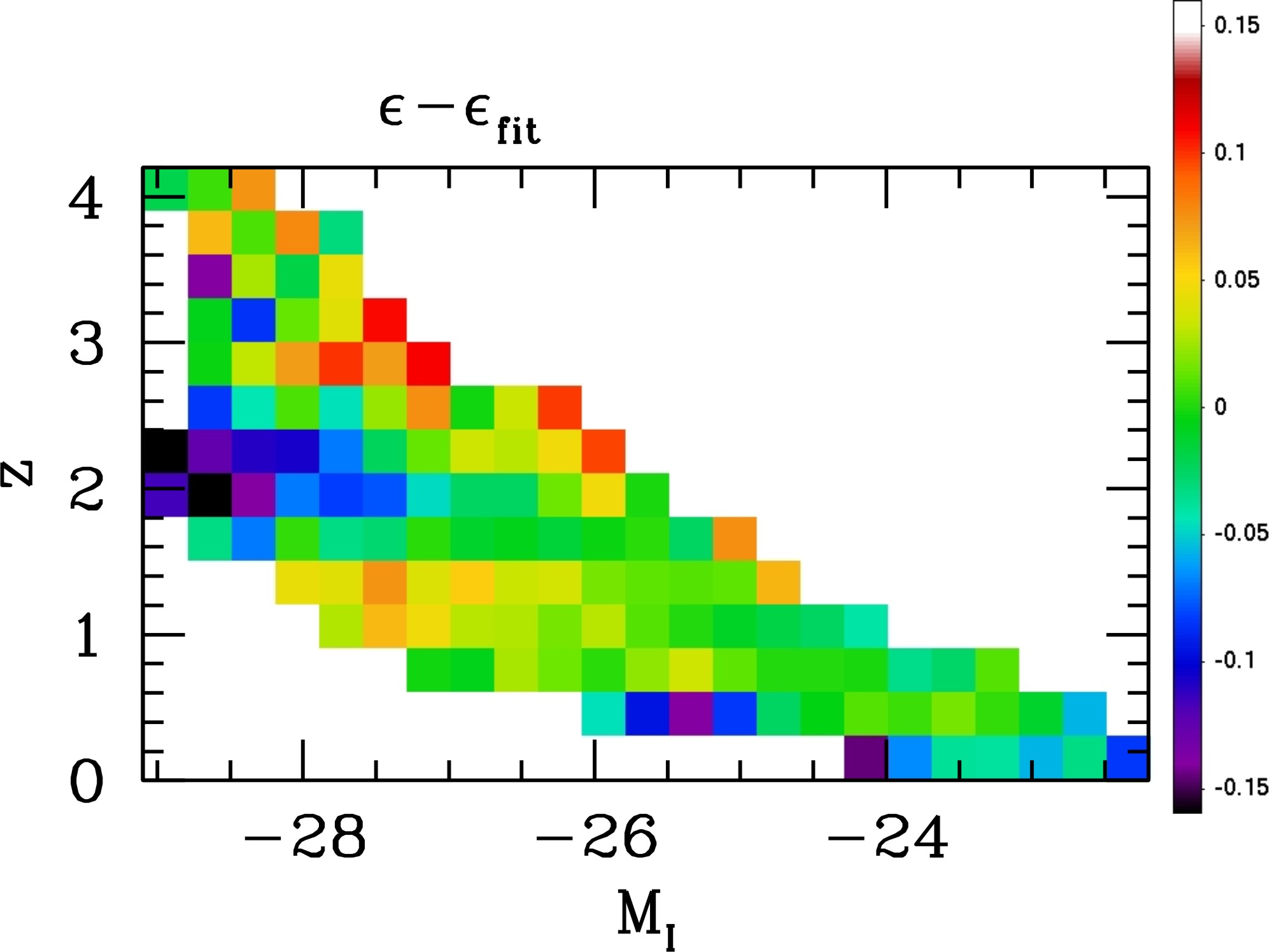}}
\par \centering}
\caption{$Top$ Average of $\log _{10}\epsilon$ (logarithm of the Eddington ratio) 
in bins of $\Delta z=0.3$ in redshift and
$\Delta M_i=0.3$ in the $i$-rest absolute magnitude. ($Bottom$) Difference between the data and the best fit model.}
\label{Fig:edd}
\end{figure}

Table 1 presents the values of the fits for $r=\log _{10}M_{BH}$ and
$r=\log _{10}\epsilon$. Figures \ref{Fig:mass} and \ref{Fig:edd} show the
average values in bins of redshift and absolute magnitude of these masses and
Eddington ratios and the differences between the data and the fit. Only bins enclosing
at least 10 objects have been considered in these plots. The  maximum difference 
between the data and the model is 0.15 dex with no clear dependence of these
differences on redshift or magnitude. These mean differences per bin are 0.067 for
both the mass and the Eddington ratio fits. So, we conclude that our simple bi-linear
models are good description of the data.

As said above, we do not claim here that masses (or Eddington ratios) only depend on the luminosities and the redshifts. There may be dependence on other parameters. We are analyzing $\langle M_{BH}\rangle $ in bins of fixed luminosities and redshifts; we do not analyze the exact value of $M_{BH}$ for each QSO. For a given luminosity and a given redshift, the full range of velocities is observed (within $FWHM>2000$ km/s), because the constraint of 5$\sigma $ in SNR only affects to the continuum of the line which is not dependent on the velocity, there is not a bias due to a truncation in the velocity range.

The masses determined with H$_\beta $ and MgII-2798\AA \ agree quite well each other (McLure \& Dunlop 2004, S08). The results shown in Table 1 indicate that the
coefficients of the fits for the ranges $z\le 0.7$ and $0.7<z\le 1.9$ are quite
similar. Masses at redshifts $z\ge 1.9$ were determined from  CIV-1549\AA \ lines,
and have a large uncertainty as has been shown by several authors (Vestergaard \&
Peterson 2006, S08, Netzer 2010). However, we do not find within that range of
redshift any dramatic change in the general trend of the fits when objects at $z>1.9$
are included, and then we concluded that it is possible to use the masses statistically determined from CIV. Anyway, because of the given reasons, 
results at $z\le 1.9$ are more robust.

As expected for a fixed redshift, there is a strong dependence of both masses and Eddington ratios on luminosity. There is no
surprise in that dependence because it is   
explicit in the virial theorem (see Eq. (\ref{mbh_beta_green}) for another formulation of it), which 
we used to derive mass through the   
direct relationship with luminosity, and is also implicit in the   
possible dependence of velocities on luminosity (Fine et al. 2008, 2010 do
not find such dependence).
We got an average relationship $M_{BH}\propto
L_{i,rest}^{0.605\pm 0.005}$ (derived from $c_1$ in Table 1 for $z\le
1.9$). Since roughly $L_{i,rest}$ is proportional to $L_{bol.}$\footnote{We  must bear in mind that $L_{bol.}$ in S08 is given by a linear relationship with the
continuum luminosity at a given wavelength, and since the average rest-colors of QSOs
do not change significantly with redshift, the proportionality of different
luminosities is expected.} it is also logical that the result we get is $\epsilon \propto
L_{I,rest}^{0.362\pm 0.005}$ (derived from $c_2$ in Table 1 for $z\le 1.9$); that is,
the most massive black holes do not accrete at their Eddington luminosity, but rather
all fall well shorter (Steinhardt \& Elvis 2010).

The most interesting result here is that both $M_{BH}$ and $\epsilon $ do not 
significantly depend on the redshift for a fixed luminosity, and there is not circularity in
this result. This is very well illustrated in Fig. \ref{Fig:mass}, where the change in color (representative of masses)
is produced in the horizontal direction (change of luminosities) but not in the vertical direction (change of redshifts).
There is not circularity because there are no reasons to think a priori that the rotation speeds given by the 
widths of the broad lines are dependent on or independent of the redshift, and the width of the broad lines is 
the only variable in which such a dependence could arise (see, for instance, Eq. (\ref{mbh_beta_green}])).
From the values of $b_1$, $b_2$ for 
$z\le 1.9$, we get that 
\begin{equation}
\left\langle \frac{\partial M_{BH}}{\partial z}\right\rangle=(0.138\pm 0.012)M_{BH}
,\end{equation}
\begin{equation}
\left\langle \frac{\partial \epsilon}{\partial z}\right\rangle=(-0.154\pm 0.012)\epsilon
.\end{equation}
The errors only stand for the statistical errors, not for the systematic errors which,
as shown in \S \ref{.comp}, might be of the order of the given signal. This
means that the variation of mass for QSOs of the same luminosity is around 14\% per
unit redshift, and the variation of the Eddington ratio is around -15\% per unit
redshift. This is quite small taking into account that the masses of the observed
$z\approx 2$ SDSS QSOs are on average 10 times more massive and have four times higher Eddington ratios than at $z=0.2$ (McLure \& Dunlop 2004). So the conclusion is clear: the
variation of mass and Eddington ratio at high redshift with respect to low redshift is
due mainly to selection effects in luminosity. If we could observe QSOs at high-z as faint as low-z QSOs, then they would have roughly the same mass and the same Eddington ratio than the low-z ones.

Steinhardt \& Elvis (2010) also considered this non-evolution: they found that the maximum
luminosity of the QSOs is given by a sub-Eddington limit of $L=\alpha M*L_0$ and the
slope $\alpha (z)=(0.41\pm 0.09)+(0.12\pm 0.08)z$, that is compatible with no evolution.

Croom (2011) also analyzed the black hole masses obtained by S08 data, although with a different
method: using a Kolmogorov-Smirnov test to compare the distributions of $M_{BH}$ for different redshifts and 
luminosities with those with randomized emission line velocities. 
Croom (2011) concludes that broad-line widths do not
have a significant impact on the estimation of black hole mass. This is the same thing as saying
that the mass of the black hole only depends on its luminosity, which is in agreement with
our results. Implicitly, Croom (2011) or previously S08 already gave some information which could
lead to our results. However, these authors were not explicit about the consequences in terms of
a ``non-evolution.'' Indeed, Croom (2011) did not mention anything about this non-evolution 
of the QSOs which we believe to be an important point to be discussed as a separate topic. The
aim by Croom (2011) is that black hole masses only depend on their luminosities, whereas our aim
remarks that black hole masses do not depend on redshift (for a constant luminosity). Apart from
the different method of the double-linear fit, which can be taken as an independent confirmation of Croom (2011)
result, we also extend our analysis to the systematic errors (see \S \ \ref{.comp}) and, rather than
discussing the luminosity dependence, we are herein more interested in discussing the non-dependence on
the redshift.

On average for $z\le 5$ (the first row in Table 1; neglecting the dependence on redshift, taking the value for the average redshift $z=1.35$): 
$\epsilon \approx 0.069\times 0.738^{(M_i+23)}$ ($M_i$ in AB magnitude). 
Taking into account that $L_{bol}=6.8\times 10^{36}\times 10^{-0.376M_B}$ erg/s (McLure \& Dunlop 2004) ($M_B$ in Vega magnitude), and an average color $\langle i(AB)-B(Vega)\rangle \approx 0.2$
(this comes from an average spectrum of a quasar in optical being 
$F_\nu \propto \nu ^{-0.5}$ [Richards et al. 2001], and the difference $B(Vega)=B(AB)+0.16$), 
we get

\begin{equation}
\epsilon \approx 0.22\left(\frac{L_{bol}}{10^{47}\ erg/s}\right)^{0.35}
\end{equation}
\begin{equation}
\left(\frac{M_{BH}}{10^9\ M_\odot}\right)\approx 3.4\left(\frac{L_{bol}}{10^{47}\ {\rm erg/s}}\right)^{0.65}
,\end{equation}

\begin{table*}
\caption{Bilinear fit of $\log _{10}M_{BH}=a_1+b_1z+c_1(M_i+23)$ and 
$\log _{10}\epsilon =a_2+b_2z+c_2(M_i+23)$ of the S08 sample. 
The first column indicates the redshift range used, and in brackets the corresponding
line used for the determination of masses and bolometric luminosity. 
$N$ is the number of QSOs. The quoted uncertainties are the statistical 1$\sigma$ errors.}
\begin{center}
\begin{tabular}{crcccccc}
\label{Tab:fit}
{\tiny Range of $z$} & $N$ & $a_1$ & $b_1$ & $c_1$ & $a_2$ & $b_2$ & $c_2$ \\ \hline
{\tiny All ($z\le 5.0$)} & {\tiny 49213} & {\tiny $8.208\pm 0.003$} & {\tiny $+0.052\pm 0.004$} & {\tiny $-0.250\pm 0.002$} & {\tiny $-1.103\pm 0.003$} & {\tiny $-0.042\pm 0.004$} & {\tiny $-0.132\pm 0.002$} \\	      
{\tiny $z\le 0.7$} & {\tiny 8417} & {\tiny $8.289\pm 0.017$} & {\tiny $-0.097\pm 0.044$} & {\tiny $-0.266\pm 0.008$} &
                   {\tiny  $-1.184\pm 0.017$} & {\tiny $+0.147\pm 0.043$} & {\tiny $-0.100\pm 0.008$} \\		    
{\tiny $0.7<z\le 1.9$}  & {\tiny 32738} & {\tiny $8.153\pm 0.006$} & {\tiny $+0.082\pm 0.006$} & {\tiny $-0.251\pm 0.002$} &
             {\tiny  $-1.069\pm 0.006$} & {\tiny $-0.093\pm 0.006$} & {\tiny $-0.146\pm 0.002$} \\	      
{\tiny $1.9<z\le 5.0$}  & {\tiny 8058} & {\tiny $8.176\pm 0.023$} & {\tiny $-0.067\pm 0.009$} & {\tiny $-0.326\pm 0.007$} &
             {\tiny  $-1.050\pm 0.023$} & {\tiny $+0.095\pm 0.009$} & {\tiny $-0.042\pm 0.007$} \\	           	      	      
{\tiny $z\le 1.9$} & {\tiny 41155} & {\tiny $8.216\pm 0.004$} & {\tiny $+0.060\pm 0.006$} & {\tiny $-0.242\pm 0.002$} &
             {\tiny  $-1.105\pm 0.004$} & {\tiny $-0.067\pm 0.006$} & {\tiny $-0.145\pm 0.002$} \\ 
\end{tabular}
\end{center}
\end{table*}	      

\section{Complementary analyses}
\label{.comp}

Here we present a similar analysis by using our own estimations of masses and
luminosities. We also discuss the reliability and uncertainties of the different
methods used to estimate the black hole masses and luminosities and how they affect
the results presented in the previous section. We used the sample of QSOs in SDSS-DR7
(Abazajian et al. 2009) fitting the continuum and the approppriate emission lines
following the method outlined in Guti\'errez \& L\'opez-Corredoira (2010). We used
H$_\alpha $ (only for $z<0.333$),  H$_\beta $ and [OIII]-5007\AA , and the continuum
for QSOs with $z<0.787$, $M_B<-23$, and $FWHM>2000$ km/s. We used the absolute magnitude
in B derived from the interpolation of the different filters, rather than the $i$-band
absolute magnitude with K-corrections. Only cases in which the heights of
[OIII]-5007\AA \ and H$_\beta $-broad lines were more than 3$\sigma $ over the continuum,
and  those of H$_\alpha $-broad lines were more than 5$\sigma $ over the continuum were chosen. The
number of QSOs selected was 4954. Our sample is smaller than the one by S08 at $z<0.7$
because we took a brighter constraint in brightness ($M_B<-23$ instead of $M_i<-22$). 

{\tiny

\begin{table*}
\caption{Bilinear fit of $\log _{10}M_{BH}=a_1+b_1z+c_1(M_B+23)$ and 
$\log _{10}\epsilon =a_2+b_2z+c_2(M_B+23)$ of our sample (see \S \ref{.comp}),
and the estimators of mass and bolometric luminosity given in \S \ref{.comp}.
The first column indicates the redshift range and the
line/cont. used for the determination of the bolometric luminosity; c.e. indicates
correction of total extinction for the flux of H$_\beta $.
$N$ is the number of QSOs. The given errors are only statistical values.}
\begin{center}
\begin{tabular}{cccccccc}
{\tiny Lines} & $N$ & $a_1$ & $b_1$ & $c_1$ & $a_2$ & $b_2$ & $c_2$ \\ \hline
{\tiny $z<0.787$/OIII} & {\tiny 4954} &  {\tiny $8.43\pm 0.02$} & {\tiny $-0.23\pm 0.04$}  & {\tiny $-0.293\pm 0.008$} &
		{\tiny $-0.80\pm 0.03$} & {\tiny $0.20\pm 0.04$}  & {\tiny $-0.055\pm 0.010$} \\
{\tiny $z<0.787$/M$_B$} &  " &" " & " " & " " &
		{\tiny $-1.06\pm 0.02$} & {\tiny $0.23\pm 0.04$}  & {\tiny $-0.083\pm 0.008$} \\
{\tiny $z<0.787$/Cont.5100} &  " &  " " & " " & " " &
		{\tiny $-1.10\pm 0.02$} & {\tiny $0.06\pm 0.04$}  & {\tiny $-0.095\pm 0.007$} \\
{\tiny $z<0.333$/OIII}    & {\tiny 218} & {\tiny $8.30\pm 0.10$} & {\tiny $-0.13\pm 0.35$}   & {\tiny $-0.31\pm 0.04$} &
		{\tiny $-0.58\pm 0.15$}   & {\tiny $-0.07\pm 0.55$}   & {\tiny $-0.03\pm 0.06$} \\
{\tiny $z<0.333$;c.e./OIII} & " & {\tiny $8.25\pm 0.12$}   & {\tiny $0.01\pm 0.45$}    & {\tiny $-0.29\pm 0.05$} &
		{\tiny $-0.53\pm 0.15$}   & {\tiny $-0.22\pm 0.55$}   & {\tiny $-0.04\pm 0.06$} \\
\end{tabular}
\end{center}
\end{table*}	   

}

\subsection{Estimators of black hole masses}

There are several ways to calculate black hole masses (McGill et al. 2008), which basically differ in the lines used and the way to estimate the size of the broad line regions: either
the luminosity of a line or the luminosity of the continuum. For the broad-line region velocity
there is a general consensus that it is obtained from the square of the width of some broad line.
The different methods can differ by up to $0.38\pm 0.05$ dex in the mean,
or $0.13\pm 0.05$ dex, if the same virial coefficient is adopted (McGill et al. 2008).
This error mainly affects the calibration of the coefficients $a_1$, $a_2$ given 
in Table 1, but also the coefficients $b_1$, $b_2$.
In order to illustrate this fact, we have carried out calculations of masses with an estimator different from S08, using the flux of H$_\beta $ instead of the flux
of the continuum to derive the size of the broad line region (Green \& Ho 2005)

\begin{equation}
M_{BH}(H_\beta )=1.5\times 10^8\left(\frac{L _{H_\beta }}{3.8\times 10^{43}\ {\rm erg/s}}
\right)^{0.56}
\end{equation}\[
\left(\frac{\sigma _{H_\beta }}{1000\ {\rm km/s}}
\right)^{2.00}\ {\rm M_\odot}
\label{mbh_beta_green}
.\]
The result, as given in Table 2, is an $a_1$ around 0.15 dex larger than that with the S08 
sample at $z<0.7$, and $b_1$ around 0.15 dex smaller, so we must 
conclude that the systematic errors associated with the mass estimators are of 
this order in $a_1$ and $b_1$, 0.1-0.2 dex, much larger than the statistical errors. Note that, using this method, since we truncate the sample of QSOs with low height of the H$_\beta $ line (which means very high velocity), for a given luminosity of the line, we might remove some very high mass candidates, and introduce some bias. However, the aim here was checking the robustness of the mass estimator, and the ``unbiased'' statistics are only carried out with the method in Section \ref{.samplemethod}.

\subsection{Estimators of bolometric luminosities}

Here we will compare three different methods, and the differences between them will give
us an estimation of the systematic errors associated with that calculation.

\begin{itemize}

\item Using the luminosity of the [OIII]-5007\AA \ line ($L_{OIII,5007}$; corrected for Galactic
extinction) (Heckman et al. 2004):

\begin{equation}
L_{bol_1}=3500\ L_{OIII,5007}	
\end{equation}	
We assume for QSOs that the luminosity of OIII lines due to star formation is negligible.
There is however another more important correction to carry out.
Due to the small diameter of the SDSS fiber (3"), we may lose some small amount
of light if the OIII region is somewhat spread. According to Bennert et al. (2002), the linear
radius of the OIII regions is:
\begin{equation}
R_{OIII}\approx 3.2\times 10^{-19}\left(\frac{L_{\rm OIII,5007}}{erg s^{-1}}\right)^{0.52}\ {\rm pc}
,\end{equation}
and with a surface brightness having radial dependence $L(r)\propto r^\delta $ (Bennert et al. 
2006) and an average $\delta =-2.95$, we get that at half of $R_{OIII}$, it is equivalent
to a Gaussian distribution with $\sigma \sim
0.3 R_{OIII}$. This $\sigma $ must be quadratically added 
to the seeing for OIII lines, while the continuum covers a much reduced area and its spread only stems from the seeing. For a 
Gaussian distribution of light, the amount of lost light outside a fiber of radius $\theta _f$
(=1.5" for SDSS) will be
$
Lost_{\sigma }=e^{-\frac{1}{2}\left(\frac{\theta _f}{\theta _\sigma}\right)^2}
$
where $\theta _\sigma $ is the angular size corresponding to the linear size 
of $\sigma $ (using the standard cosmological parameters). The corrected luminosity will
be $L_{OIII,5007}^{corr.}=\frac{L_{OIII,5007}}{1-Lost_{\sigma -OIII}
+Lost_{\sigma -cont.}}$.
This correction is small ($<5$\%) in most of the cases but more significant in some sources, in particular at low z.

There might also be some dependence on this relationship with the type of AGN (Netzer 2010).
We assume that the difference from the Seyfert 1 used by Heckman et al. (2004) and QSOs 
is negligible. A more precise
estimator would be obtained 
using both [OIII] and [OI] lines (Netzer 2010), but [OI] would be more
affected by star formation contamination and lost of light outside the fiber due to its
higher spread.
	
\item Using the absolute magnitude in the $B$-band with K-correction and correction for Galactic
extinction, $M_B$ (Vega calibrated) (McLure \& Dunlop 2004, eq. (C1))

\begin{equation}
L_{bol_2}=6.8\times 10^{36}10^{-0.376M_B}\ {\rm erg/s}
\end{equation}	

\item Using the continuum luminosity at 5100 \AA \ (McLure \& Dunlop 2004, \S C1)

\begin{equation}
L_{bol_3}=9.8[\lambda L_{cont.\lambda}](5100\ \AA)
,\end{equation}	
where $L_{cont.\lambda }$ is the luminosity per 
unit wavelength at a given $\lambda $ at rest, and includes Galactic extinction correction
as well. This is the method used by S08, although with their own calibration.

\end{itemize}

With the three different methods, we get values of $b_2$ equal to 0.20, 0.23 and 0.06 respectively,
to be compared with 0.15 for the S08 subsample at $z<0.7$. Again we see that the uncertainties due
to the use of different estimators are on the order of 0.1-0.2 dex in the coefficient $b_2$.

\subsection{Extinction}

Apart from the Galactic extinction, which is easy to correct, there might be some
intergalactic extinction,  extinction from the host galaxy of the QSO, and extinction
from the torus of the QSO itself. These would affect the measured fluxes,
either for the lines or the continuum. We will explore herein the relevance and
consequence of this effect.

The ratio of the broad lines $H_\alpha $, $H_\beta $
can be used to derive the extinction in each galaxy; the luminosities corrected
for extinction (using Dessauges-Zavadsky et al. 2000, which assumed 
a ratio of 3.1 for AGNs, and that the variations of this value are due 
only to extinction ) would be
\begin{equation}
L_{H_\alpha }^{c.e.}=\frac{L _{H_\alpha }^3}{3.1^2L _{H_\beta }^2}, \hspace{1cm} L_{H_\beta }^{c.e.}=\frac{L _{H_\alpha }^{c.e.}}{3.1}
,\end{equation}
Due to the spectral coverage of SDSS spectra, $H_\alpha $ is only available within $z<0.333$, which means there are 218 QSOs in our sample.
La Mura et al. (2007) also used Balmer lines AGNs with z < 0.4
in their analysis.
In this subsample, we can carry out the calculation of the masses with the extinction corrected fluxes. The results are presented in Table 2. Here, due to the low number of QSOs, the
statistical errors are larger than the effect of the extinction correction. Anyway,
apart from the statistical errors (which are equal with or without extinction correction),
we can see that the extinction effect reduces by 0.1-0.2 dex the value of the 
$b_1$ and increases $b_2$ by the same amount.

A bilinear fit of the ratio of both Balmer lines gives
\begin{equation}
\log _{10}\frac{F_{H_\alpha }}{3.1F_{H_\beta }}=
(-0.035\pm 0.051)+(0.11\pm 0.19)z
\end{equation}\[
+(0.005\pm 0.022)(M_B+23)
,\]
so no significant dependence on redshift is found and then  no significant
detection of intergalactic extinction (which should be increasing with $z$) was found.

\section{Conclusions and discussion}

The main conclusion of our analysis is that both the mass and the Eddington ratio of
the black holes for a QSO with a given luminosity do not evolve with redshift. Or in other words,
the luminosity of a QSO does not evolve with redshift for a given mass. More
precissely and considering  systematic uncertainties  $\sim 0.2-0.3$ dex in the
estimation of masses and luminosities, we conclude that the evolution in redshift, if
any, is very small compared  to the change in mean luminosity of the population of
QSOs at low redshift with respect to such a population at high redshifts. This implies 
an important result on the nature of QSOs, i.e. local QSOs are
intrinsically less massive than QSOs at high redshift. 
Labita et al. (2009a) derived that
the maximum mass of a black hole in a QSO is a function of the redshift: $\log
_{10}M_{BH}=(0.34z+8.99)$ M$_\odot$ up to redshift 1.9, or proportional to
$(1+z)^{1.64}$ if extended up to a redshift of 4 (Labita et al. 2009b).
This lack of the signature of active massive AGN black holes in the local Universe cannot be related with a
possible decline in the rate of formation of QSOs (this would affect the density of
QSOs but not their average mass; and indeed there is evidence for the change in the comoving
density of QSOs of a given mass; Steinhardt \& Elvis 2011, fig. 3), but because of some mechanism for the formation of huge black holes which took place in the past in the Universe, which is absent in the present Universe.

NOTE: Do not confuse the non-evolution of the black hole mass-luminosity ratio (the result of 
this paper) with the non-evolution of mechanisms which produce such black holes. Evidently,
as said in the introduction, some evolution in the birth of new QSOs must take place in order
to explain the absence of very bright QSOs at low redshift. 

The fact that the Eddington ratio of the QSO does not change with $z$ for a given
luminosity/mass counters scenarios which explain the evolution of the
luminosity of QSOs in terms of the change in the environment of the AGNs, like variations
in the accretion rate. Our result is at odds with the common assumption that relates
the influence  of companion galaxies with the mechanism of feeding the  black hole of
the QSO (Stockton 1982; Canalizo \& Stockton 2001). Horst \& Duschl (2008) presented
the results of a simple cosmological model combined with an evolutionary scenario in
which both the formation of the black hole as well as the gas accretion  onto it are
triggered by major mergers of gas-rich galaxies. Despite the very  generous number of
approximations, their model reproduces the quasar density evolution in  remarkable
agreement with some observations. 
However, we do not see this decrease of gas
accretion here, so the environment does not seem to be the major factor responsible for the change in luminosity of QSOs. We do not deny, however, the effect of the environment on the mechanism of formation/turn-off of its own massive black hole, although the synchronization
of all very-massive QSOs in an epoch turning off nearly at the same time is not understood, because
galaxies continue to merge and virialize at some rate at later epochs (Steinhardt \& Elvis 2011).

Other results from other papers are also apparently at odds with the idea
of powerful AGNs in a rich highly interactive environment. 
Coldwell \& Lambas (2006) showed that 
quasars at $z<0.2$ systematically avoid high density regions,
living in regions less dense than cluster environments.
At $0.5\le z\le 0.8$, only 10\% of QSOs live in relatively rich clusters, 
and 45\% of them in field-like environments (Wold et al. 2001). 
We might also consider that galaxies in rich clusters are stripped of their 
interstellar medium by harassment, so it would be reasonable that the QSO activity 
is less than in the field galaxies, but the ratio of spiral galaxies with 
non-stripped gas is still high enough to consider that there should be activity being triggered.
More recently, Cisternas et al. (2011) showed directly that
there is not an enhanced frequency of merger signatures for the AGN hosts
with respect to other galaxies, so this points out that mergers should not
be an important element for the triggering of activity.

There are other kind of models on the origins and the early evolution of QSOs and 
supermassive black holes (review at Djorgovski et al. 2008). Many works have related 
the evolution of QSOs with their star formation ratios; e. g.  Haiman et al. (2007)
assume that star formation in spheroids (elliptical galaxies and bulges of late-type
galaxies) and black hole fueling are proportional to one another at all times, and
fitting conveniently some parameters get a model of luminosity evolution of QSOs in
agreement with observed data with quasar lifetimes $\approx 6-8\times 10^7$ yr, without
a compelling need for any of the model parameters to evolve with redshift between
$0<z<6$. This result supports the direct connection between the build up of spheroids
and their nuclear supermassive-black-holes.

The existence of very massive black holes is only at high $z$. Perhaps
it could be related with the higher ratio of mergers and then star formation in the
past. We wonder whether it might have something to do with the excess of very massive
galaxies at high-$z$, which is still not completely understood within semianalytical
hierarchical $\Lambda $CDM models (e.g., Fontana et al. 2009).  Indeed, the mass of
the black hole has remained proportional to the stellar mass of their host galaxies for at
least the last 9 Gyr (Jahnke et al. 2009). Or perhaps it has something to do with
the larger  average density of the Universe, or the angular momentum of some
components of the galaxies (at high $z$, the black holes rotate faster; Netzer 2010).
However, where are the black holes with masses larger than $10^{10}$ M$_\odot $ which were
frequent in the past? We do not know any mechanism by which black holes
can reduce its mass. Also, one should
not lose sight of some solutions in which, for some reason (e.g., radiation emitted
in beams rather than isotropically over $4\pi$ stereoradians, non-cosmological
redshifts, wrong cosmological model, etc.), the luminosity of the QSOs does not correspond
to $4\pi d_L(z)^2Flux$, with $d_L(z)$ the luminosity distance given by standard 
cosmology, and consequently both the luminosities and the masses\footnote{The
calculation of the mass depends on the luminosity of the continuum, like in the
application by S08, or of some line, like in Eq. (\ref{mbh_beta_green}). Note,
moreover, that there is not a  significant variation of the width of broad lines with
redshift (S08, fig. 3) so this  reinforces the idea that huge masses are obtained
because huge luminosities are measured, and an overestimation of the luminosity
directly produces an overestimation of the masses.}  would be erroneously determined.
Clearly the question remains open and it is beyond the scope of the present
paper. 

\normalem
\begin{acknowledgements}
Thanks are given to J. A. Acosta-Pulido (IAC, Tenerife, Spain) and the anonymous referee in RAA 
for helpful comments on this paper. 
Funding for the creation and distribution of the SDSS Archive has been provided by the
Alfred P. Sloan Foundation, the Participating  Institutions, the National Aeronautics
and Space Administration, the  National Science Foundation, the U.S. Department of
Energy, the Japanese  Monbukagakusho, and the Max Planck Society. The SDSS Web site
is  http://www.sdss.org/. The SDSS is managed by the Astrophysical Research 
Consortium (ARC) for the Participating Institutions. The Participating  Institutions
are The University of Chicago, Fermilab, the Institute for  Advanced Study, the Japan
Participation Group, the Johns Hopkins University,  the Korean Scientist Group, Los
Alamos National Laboratory, the Max-Planck-Institute for Astronomy (MPIA), the
Max-Planck-Institute for  Astrophysics (MPA), New Mexico State University, University
of Pittsburgh, University of Portsmouth, Princeton University, the United States
Naval Observatory, and the University of Washington.
\end{acknowledgements}

\label{lastpage}

\end{document}